# Non-volatile hybrid optical phase shifter driven by a ferroelectric transistor

Rui Tang[1,2], Kouhei Watanabe[1,2], Masahiro Fujita[1], Hanzhi Tang[1], Tomohiro Akazawa[1], Kasidit Toprasertpong[1], Shinichi Takagi[1], and Mitsuru Takenaka[1*]

[1]Department of Electrical Engineering and Information Systems, The University of Tokyo, Tokyo 113-8656, Japan
[2]These authors contributed equally
[*]Email: takenaka@mosfet.t.u-tokyo.ac.jp

**Abstract**: Optical phase shifters are essential elements in photonic integrated circuits (PICs) and function as a direct interface to program the PICs. Non-volatile phase shifters, which can retain information without a power supply, are highly desirable for low-power static operations. Here a non-volatile optical phase shifter is demonstrated by driving a III-V/Si hybrid metal-oxide-semiconductor (MOS) phase shifter with a ferroelectric field-effect transistor (FeFET) operating in the source follower mode. Owing to the various polarization states in the FeFET, multistate non-volatile phase shifts up to $1.25\pi$ are obtained with CMOS-compatible operation voltages and low switching energy up to 3.3 nJ. Furthermore, a crossbar array architecture is proposed to simplify the control of non-volatile phase shifters in large-scale PICs and its feasibility is verified by confirming the selective write-in operation of a targeted FeFET with a negligible disturbance to the others. This work paves the way for realizing large-scale non-volatile programmable PICs for emerging computing applications such as deep learning and quantum computing.

**Introduction**

As Si-based complementary metal-oxide-semiconductor (CMOS) electronics are approaching the physical limit of miniaturization, photonic integrated circuits (PICs) are becoming increasingly critical in computing and information processing. Photons tend to have weak interactions with each other and therefore are linear and parallel by nature, making them particularly suitable for various parallel tasks such as matrix multiplication [1–3]. Directly processing these tasks in the optical domain can significantly reduce delays and potentially lower energy consumption compared with the case of using electronics [4]. To tackle diverse tasks, PICs must be reconfigurable and programmable, which are realized by tailoring the on-chip transmission property via optical phase shifters [5]. The phase shifter is an essential component of PICs which shifts the phase of light by altering the refractive index of the optical waveguide. Further through coherent interference between the light in adjacent waveguides, phase shifters function as a direct interface to configure the PICs. Phase shifters based on various mechanisms have been realized [6–9], yet most of them are volatile: the information will be lost once the power is turned off. Non-volatile phase shifters, by contrast, can retain information without a power supply, thus significantly simplifying control complexity and reducing power consumption during static operations.

Non-volatile phase shifters can be realized in several ways. The use of low-loss phase-change materials (PCMs) such as $Sb_2Se_3$ is actively being investigated [10–24]. The phase transition of PCM provides a non-volatile change in the refractive index, but the electrical switching through Joule heating typically requires a high voltage when switching the PCM from the crystalline state to the amorphous state [17], making it incompatible with CMOS drivers. Another approach involves using ferroelectric materials such as barium titanate (BTO), integrated into the waveguide to achieve a non-volatile phase shift via the Pockels effect [25]. However, its operation also requires a high voltage of up to 12 V [26]. In addition to the need for high operation voltages, controlling a large-scale PIC with a large number of non-volatile phase shifters is challenging. Thus far, potential methods which can simplify the control are rarely discussed in previous works.

Here we demonstrate a novel non-volatile optical phase shifter by driving a III-V/Si hybrid metal-oxide-semiconductor (MOS) phase shifter with a Si ferroelectric field-effect transistor (FeFET) based on hafnium zirconium oxide (HZO) [27–31]. The voltage-driven MOS phase shifter enables efficient and ultralow-power (< 100 pW) phase modulation with a CMOS-compatible voltage via free carriers accumulated at the MOS interface [7,32]. The FeFET operates in the source follower mode [33,34], in which its output voltage is shifted by a non-volatile threshold voltage determined by the polarization state in the ferroelectric layer. By applying a voltage pulse to the gate of the FeFET to configure the threshold voltage, thereby configuring the voltage applied to the MOS phase shifter, we demonstrate multistate non-volatile phase shifts with the proposed scheme. Furthermore, we propose a crossbar array architecture to significantly simplify the control of phase shifters in large-scale PICs. We verify the feasibility of the proposed architecture by observing the responses of the threshold voltage of a FeFET in various states. This work paves the way for the realization of large-scale non-volatile programmable PICs, which can ultimately enable ultrafast and energy-efficient information processing in the optical domain.

**Results**

**Principle.** The proposed scheme is illustrated in Fig. 1. The hybrid MOS phase shifter consists of n-InGaAsP, $Al_2O_3$, and p-Si layers. If a positive voltage is applied between the n-InGaAsP and p-Si, free electrons and holes accumulate at the InGaAsP and Si MOS interfaces, respectively, and therefore alter the refractive index primarily through the plasma dispersion and band-filling effects [7,35]. Since the electron-induced refractive index change is more than ten times greater in InGaAsP than in Si, the hybrid MOS phase shifter enables efficient and low-loss phase tuning. During static operations, since the power consumption is only determined by the leakage current of the MOS capacitor, ultralow-power operations can be realized. Previously, we have demonstrated such low-loss, efficient, and ultralow-power hybrid MOS phase shifters [7,32,36,37]. The FeFET contains a thin ferroelectric HZO layer as the gate dielectric, in which the polarization direction can be reoriented by an external electric field applied through the gate terminal and retained afterward [27]. The HZO-based FeFET is compatible with the CMOS process and can exhibit strong ferroelectricity at room temperature even with a several-nanometers-thick HZO layer [30]. Depending on the polarization state, the threshold voltage of the FeFET is continuously tunable and stored in a non-volatile manner. For a FeFET driving a capacitor in the source follower mode [33], the output voltage ($V_o$) is approximately given by

$$V_o = \begin{cases} 0 & (V_g < V_{th}) \\ V_g - V_{th} & (V_{th} < V_g < V_i + V_{th}), \\ V_i & (V_g > V_i + V_{th}) \end{cases} \quad (1)$$

where $V_{th}$, $V_g$, and $V_i$ are the threshold voltage, gate voltage, and input voltage, respectively (see Fig. 1(b)). For a fixed $V_g$, the output voltage can be tuned by adjusting the non-volatile threshold voltage. Therefore, by driving the MOS phase shifter with a FeFET operating in the source follower mode, non-volatile phase shifts can be obtained. The attainable range of the phase shift can be adjusted by engineering the memory window of the FeFET [38,39], which determines the tunable range of $V_o$, and/or by engineering the modulation efficiency and length of the MOS phase shifter. In this work, MOS phase shifters of various lengths and FeFETs of various sizes are fabricated separately on different chips for proof-of-concept demonstration purposes, as shown in Fig. 2.

**Non-volatile operations.** We demonstrate non-volatile phase shifting operations by driving a 1.5-mm-long III-V/Si hybrid MOS phase shifter on one arm of an asymmetric Mach-Zehnder interferometer (AMZI) with an HZO-based FeFET (gate length/width: 25/100 µm). The 1.5-mm-long phase shifter is chosen to obtain a large phase shift since the induced phase shift is proportional to the device length. Although large FeFETs are intentionally chosen for easier in-house fabrication, there are no factors hindering the miniaturization of FeFETs if they are fabricated in mature CMOS manufacturing lines. The typical characteristic of our fabricated FeFETs is provided in the supplementary. The AMZI is made from half-etched Si rib waveguide with a slab height of 110 nm. The path length difference between the long and short arms is 20 µm. The measured static insertion loss and dynamic modulation loss of the MOS phase shifter are 2.57 dB and 0.31 dB/π, respectively. The static loss can be significantly reduced by improving the design of tapers inside the phase shifter (Details are provided in the supplementary). The power consumptions of the MOS phase shifter for $\pi$ and $2\pi$ phase shifts are 13.3 pW and 96.3 pW, respectively. Because the leakage current in the phase shifter increases at a higher voltage, the power consumption is not linearly proportional to the phase shift.

The source follower operation of the FeFET is first characterized, as shown in Fig. 3(a). After applying a negative reset pulse (-3 ~ -2 V, 10 ms) to the gate, the FeFET exhibits a high threshold voltage, resulting in a low $V_o$; after applying a positive voltage pulse to the gate, the threshold voltage decreases, resulting in a high $V_o$ for the same $V_g$. As shown in Fig. 3(a), $V_o$ is tuned by 0.78 V between the low $V_{th}$ and high $V_{th}$ states at a $V_g$ of 0.8 V. A larger tunable range of $V_o$ can be obtained by increasing the memory window of the FeFET. Previously, an HZO-based FeFET with a large memory window of 3.12 V has been demonstrated [40].

The source terminal of the FeFET is then electrically connected to the MOS phase shifter, and the transmission spectra of the AMZI are measured when $V_g$ and $V_b$ (see Fig. 1(b)) are 0.8 V and -0.5 V, respectively. A negative bias ($V_b$) is applied to the MOS phase shifter to preset it into the accumulation state. A clear shift of the AMZI spectrum is confirmed, as shown in Fig. 3(b), indicating that an optical phase shift is induced by the change in $V_o$. Here, a decrease in the optical power at the low $V_{th}$ state is primarily caused by the drift of optical fibers edge-coupled to Si waveguides. Since the FeFET operates with the same $V_g$ and $V_i$ between the low $V_{th}$ and high $V_{th}$ states, a non-volatile optical phase shifter is successfully achieved. We further apply pulses with increasing voltage from 2 to 3 V to the gate after a reset pulse (Fig. 3(c)), and then measure $V_o$ and extract the phase shift from the spectra after each pulse. As the polarization in the HZO film gradually changes with the increasing voltage pulses, multistate $V_o$ is obtained, as shown in Fig. 3(d). Since the phase shift is almost linearly proportional to $V_o$, as shown in the inset of Fig. 3(d), multistate phase shifts up to $1.25\pi$ are successfully demonstrated. Meanwhile, it can be seen that $V_o$ only changes slightly for pulse voltages within the range of 2.0 ~ 2.2 V. Therefore, we use 2 V pulses to set the FeFET into the initial state in the following experiments. Multiple FeFETs of different sizes are also characterized. While the threshold voltages and memory windows are not exactly the same, electrical responses similar to that shown in Fig. 3(d) are obtained. These results are provided in the supplementary.

**Switching characteristics.** The energy consumption of FeFET during write-in and reset operations can be characterized from polarization-voltage (P-V) measurements, following the method described in Ref. [41]. For 10-ms gate pulses, the measured switching energy of FeFET is shown in Fig. 4(a). An energy of up to 3.2 nJ is sufficient for write-in operations. For the reset pulse, the measured energy consumption is up to 3.3 nJ. Since the switching energy of FeFET depends on its size [41], power consumption orders of magnitude smaller can be expected if the fin field-effect transistor (FinFET) is used. In addition to pulse voltage, the shift of threshold voltage also depends on the pulse width. We apply voltage pulses ranging from 10 ms to 10 µs to the gate and then measure the shift of threshold voltage. The result is shown in Fig. 4(b). As expected, a short pulse induces a small amount of threshold shift compared with a long pulse with the same voltage. For pulse widths shorter than 1 ms, high voltages incompatible with CMOS drivers may be needed to obtain a desired phase shift. Therefore, 10 ms pulses are used throughout our experiments. The switching speed of our non-volatile phase shifter is fundamentally limited by the polarization switching speed of the FeFET. Here, since the large FeFET used in this work results in a large RC time constant, the polarization switching speed is relatively low. We expect that reducing the size of the FeFET will improve the polarization switching speed, since sub-nanosecond polarization switching time has been demonstrated in previous works [42,43].

**Crossbar array control.** For large-scale PICs, the individual control of all phase shifters has been a long-standing problem. Traditionally, $N^2$ phase shifters require $N^2$ driving channels, resulting in significant control difficulties for $N > 100$. Here, we propose a crossbar array architecture for FeFET-driven non-volatile MOS phase shifters to simplify their control in large-scale PICs, as shown in Fig. 5(a). This "2T1C" architecture, which consists of 2 transistors (one FeFET and one non-ferroelectric FET) and 1 capacitor (the MOS phase shifter), uses horizontal word/selection lines and vertical bit lines to selectively control a single phase shifter in the array. The schematic structure of a single cell in the crossbar array is illustrated in Fig. 5(b). In this architecture, the word lines and bit lines require individual controls, whereas all the selection lines only need be driven digitally by a single channel. Therefore, the required driving channels are reduced to $2N+1$. While the MOS phase shifter requires an additional bias voltage ($V_b$), this bias voltage does not require modulation and can be shared by all the phase shifters.

Figure 5(c) shows the approach to selectively reset a target FeFET in the array. A negative voltage pulse ($V_{reset}$) is applied to the corresponding word line. In order to not disturb other FeFETs connected to the same word line, all selection lines are turned on and the same $V_{reset}$ is simultaneously applied to all other bit lines. The ON-state selection lines switch on the non-ferroelectric FETs so that the source and drain terminals of each FeFET have the same electric potential. In this way, an inverse electric field is only established in the target FeFET during the reset pulse. It is important to note that a weaker electric field along the same direction does not overwrite the polarization state induced by a previous stronger electric field. When using a -2 V reset voltage, the non-target FeFET in the state 4 experiences an effective +2 V voltage pulse, which does not affect its current polarization state because the write-in voltages in this scheme are higher than 2 V, as can be seen from Fig. 3(d).

To program a target FeFET in the crossbar array, voltage pulses ($V_{wl}$, $V_{bl}$) are applied to the corresponding word line and bit line, respectively, with other word and bit lines remaining inactive, as shown in Fig. 5(d). $V_{wl}$ and $V_{bl}$ are chosen to only activate the target FeFET, without disturbing the rest. Depending on the voltages applied to each terminal, the FeFETs in the proposed architecture can experience the 4 different states shown in Fig. 5(d) and Fig. 6(a). Ideally, only the target FeFET (state 1) is affected and all other FeFETs (states 2-4) should be unaffected. Since the function of the non-ferroelectric FETs is to ensure the source and drain terminals of each FeFET have the same electric potential during write-in operations, it allows us to emulate the four states with a single FeFET by simultaneously applying various voltage pulses to the 3 terminals of the FeFET via probes. Then, the source follower operation of the FeFET is characterized and the threshold voltage is extracted. By comparing the threshold voltage before and after the voltage pulses, we can confirm whether the FeFET has been rewritten or not in these states. For state 1, we first set the FeFET to the initial state by applying a positive voltage pulse (2 V, 10 ms) to the gate and then apply various $V_{wl}$ and $V_{bl}$ pulse combinations to the FeFET. For each combination, the shift of threshold voltage with respect to the initial state is extracted and shown in Fig. 6(b). As can be seen, for voltage differences ($V_{wl}$-$V_{bl}$) beyond 2.2 V, the threshold voltage decreases as desired; for voltage differences smaller than the initial voltage pulse (2 V), the shift of threshold voltage is negligible. For state 2, we apply various $V_{wl}$ pulses ranging from 0.2 to 1.8 V to the gate and characterize the source follower operation after each pulse. All the results are plotted in Fig. 6(c), with the inset showing the shift of the threshold voltage with respect to the initial state. Similarly, Fig. 6(d) shows the results for state 3 after various $V_{bl}$ pulses ranging from -0.2 to -1.8 V. We can see that for both states, the shifts of the threshold voltage are negligible provided that the difference between $V_{wl}$ and $V_{bl}$ is smaller than the initial voltage pulse (2 V). For state 4, it is clear that no changes will occur since no voltages are applied to the FeFET. Therefore, by emulating different states that the FeFET can experience in the crossbar array, we verified that it is possible to program a single phase shifter using the proposed architecture. It is worth noting that although negative voltages are used here, a common bias can be applied to all terminals to ensure that all voltages are positive.

**Discussion and outlook.** We compare the performance of our non-volatile phase shifter with those in previous works in Table 1. The non-volatile phase shifter demonstrated in this work simultaneously enables CMOS-compatible operation voltage, low switching energy, multistate operations, and compatibility with crossbar array control. The endurance of our non-volatile phase shifter is primarily determined by the FeFET. An HZO-based FeFET with $10^{11}$ endurance cycles has been demonstrated recently [44], suggesting that our scheme could achieve a high endurance by improving the FeFET. It is noteworthy that we claim our phase shifter to be non-volatile because the information, which in our case is the threshold voltage of the FeFET, is not lost when the power supply is switched off. The phase shifter can resume its previous state once the power supply is switched on again. Meanwhile, the operation of our phase shifter requires external voltages, which is a drawback compared with other schemes. However, from a practical point of view, this may not be a serious issue since the static power consumption is determined by the leakage current in the MOS phase shifter, which is sufficiently small (sub-nanoampere level) in the demonstrated device. Compared with PCMs, the length of our device needs to be further reduced, which can be realized by improving the modulation efficiency of the MOS phase shifter, or increasing the memory window of the FeFET. In addition, the optical waveguide in the phase shifter region can be bent multiple times to reduce the chip length, as demonstrated in Ref. [45].

The static insertion loss can be reduced by optimizing the doping profile in the MOS phase shifter and the input/output III-V waveguide taper.

In future, the MOS phase shifter and FeFET may be integrated on the same chip by taking advantage of state-of-the-art electronics-photonics co-integration platforms [46]. More specifically, the Si layer of FeFET and optical waveguide can be formed on the same silicon-on-insulator (SOI) wafer. Next, the ferroelectric layer can be deposited and patterned in the transistor region. After necessary oxide depositions and surface planarizations, a III-V wafer can be bonded onto the SOI wafer to form the MOS phase shifter.

Table 1. Comparison of electrically switchable non-volatile phase shifters

| Ref. | Type | Operation voltage | Switching time | Switching energy | Device size | Multistate operation |
|---|---|---|---|---|---|---|
| [11] | PCM ($Ge_2Sb_2Se_4Te_1$) | 13 V (cryst.), 24 V (amorp.) | 50 ms (cryst.), 1 μs (amorp.) | 42.5 mJ (cryst.), 5.5 μJ (amorp.) | 30×30 μm$^{2*}$ | Not demonstrated |
| [14] | PCM ($Sb_2S_3$) | 0 ~ 1 V sweep (cryst.), 6 V (amorp.) | 1 s (cryst.), 200 ns (amorp.) | ~ 6.5 mJ (cryst.), N.A. (amorp.) | 8 μm (length) | Not demonstrated |
| [17] | PCM ($Sb_2Se_3$) | 6.2 V (cryst.), 21 V (amorp.) | 0.1 ~ 1 ms (cryst.), 800 ns (amorp.) | 38.4 μJ (cryst.), 176 nJ (amorp.) | 6 μm (length) | Yes |
| [19] | PCM ($Sb_2Se_3$) | 4 V (cryst.), 6.8 V (amorp.) | 220 μs (cryst.), 408 ns (amorp.) | 1.28 μJ (cryst.), 9.25 nJ (amorp.) | 6 μm (length) | Yes |
| [21] | PCM ($Ge_2Sb_2Se_4Te$) | 3.5 ~ 4 V (cryst.), 5 V (amorp.) | Up to 5 s (cryst.), 10 ~ 20 μs (amorp.) | N.A. | 10 μm (length) | Yes |
| [22] | PCM ($Sb_2S_3$) | 9.65 ~ 9.85 V (cryst.), 10 V (amorp.) | 550 ns | 1.02 ~ 1.06 μJ (cryst.), 1.1 μJ (amorp.) | 10 μm (length) | Yes |
| [26] | BTO-embedded waveguide | 5 ~ 12 V | 0.7 s (initialization), up to 1 s (set) | 4.6-26.7 pJ | 1 mm (length, for π-shift) | Yes |
| This work | FeFET-driven MOS phase shifter | -2 ~ -3 V (reset), 2 ~ 3 V (set) | 10 ms (reset), 10 ms (set) | ≤ 3.3 nJ (reset), ≤ 3.2 nJ (set) | 1.5 mm (length, for 1.25π-shift) | Yes |

*Free-space reflection type

**Conclusion**

We have demonstrated a non-volatile optical phase shifter by driving a III-V/Si hybrid MOS phase shifter with a FeFET operating in the source follower mode. With a 1.5-mm-long phase shifter and CMOS-compatible voltages, we achieved multistate non-volatile phase shifts up to 1.25π and low switching energies up to 3.3 nJ. Furthermore, we proposed a crossbar array architecture to simplify the control of non-volatile phase shifters in large-scale PICs and verified the feasibility by emulating various states experienced by the FeFETs. This work paves the way for the realization of large-scale non-volatile programmable PICs, which can ultimately enable ultrafast and energy-efficient information processing in the optical domain.

## Methods

**Device fabrication.** To fabricate the III-V/Si hybrid MOS phase shifter, a Si-on-insulator (SOI) wafer with a 220-nm-thick Si layer was doped by boron implantation, targeting at a p-type doping concentration of $3\times10^{17}$ cm$^{-3}$. The Si rib waveguide (rib width: 1 μm, slab thickness: 110 nm) was formed by electron-beam lithography (EBL) and inductively coupled plasma (ICP) etching. Then, a 200-nm n-doped ($5\times10^{15}$ cm$^{-3}$) In$_{0.68}$Ga$_{0.32}$As$_{0.7}$P$_{0.3}$ layer ($\lambda_g$ = 1.37 μm) was bonded to the Si waveguide using a 5-nm Al$_2$O$_3$ as the bonding interface, deposited by atomic layer deposition (ALD) at 200 ºC. The InGaAsP mesas were defined by EBL and reactive ion etching (RIE). After the deposition of SiO$_2$ cladding and via formation, a Ni/Au metal layer (50 nm/400 nm) was deposited by electron-beam (EB) evaporation and then lifted off to form contact pads. FeFETs were fabricated on a p-type (001) Si substrate with the source and drain regions doped by phosphorus ion implantation, following a similar process described in Ref. [29]. The substrate was cleaned and soaked in a HCl-H$_2$O$_2$-H$_2$O mixed solution (the SC-2 solution) to prepare 0.6-nm-thick SiO$_2$, which functions as the interfacial layer between Si and HZO. Then, 10-nm-thick ferroelectric HZO films were prepared by ALD at 300°C using tetrakis(ethylmethylamino)zirconium, tetrakis(ethylmethylamino)hafnium, and H$_2$O. The HZO layer was etched at the source/drain regions to form source/drain contacts. The remaining region of the HZO layer was left un-patterned. TiN gate electrodes were deposited by sputtering and patterned to obtain a gate width of 100 μm and various gate lengths. The devices were annealed at 400°C for 30 s to form the ferroelectric phase in HZO.

**Experimental setup.** The optical chip was manually aligned and coupled with two lensed fibers (mode diameter: 4.0 μm) for optical measurements. The coupling loss was 11.5 dB/facet, due to a non-ideal design of the edge coupler that caused a significant mode mismatch. A tunable laser (Santec, TSL-510) and an optical power meter (Santec, MPM-210H) were used to characterize the transmission spectrum of the AMZI. Electrical properties of the MOS phase shifter and FeFETs were characterized using semiconductor parameter analyzers (Keysight, 4156C & B1500A). The output voltage of the FeFET was applied to the MOS phase shifter via probes and coaxial cables.

**Data availability**

Data underlying the results presented in this paper are available from the corresponding author upon reasonable request.

**Acknowledgements**

This work is partly supported by JST, CREST Grant Numbers JPMJCR2004 and JPMJCR20C3. The authors thank Prof. K. Takeuchi and Dr. C. Matsui for the discussion about the source follower operation of a FeFET.

**Author contributions**

H.T. and K.T. fabricated the optical chip and the FeFETs, respectively. R.T., K.W., and M.F. performed the experiments. All authors contributed to preparing the manuscript. M.T. and S.T. supervised the project.

**Competing interests**

The authors declare no competing interests.

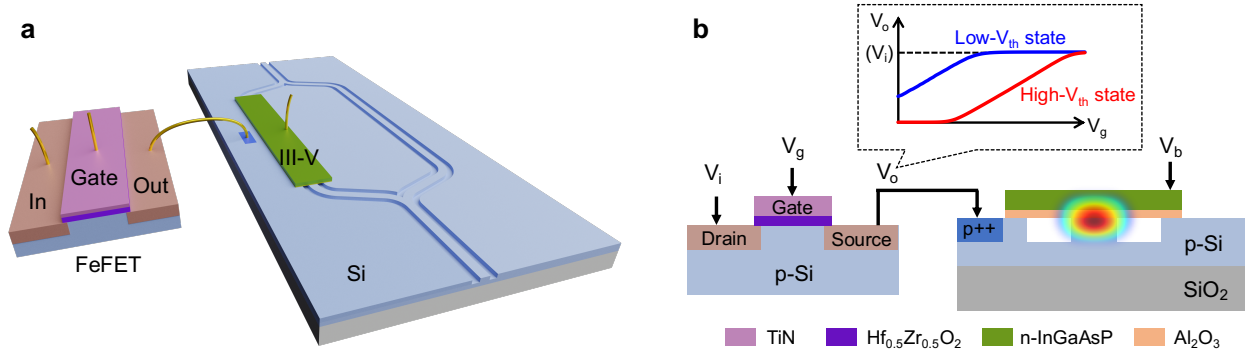

**Fig. 1. Schematic diagrams. a**. Schematic diagram of the FeFET-driven non-volatile III-V/Si hybrid MOS optical phase shifter. The drawing is not to scale. The FeFET operates in the source follower mode, in which its output voltage is shifted by a non-volatile threshold determined by the polarization state in the ferroelectric layer. The voltage-driven MOS phase shifter enables efficient and ultralow-power phase modulation via free carriers accumulated at the MOS interface. **b**. Cross-sectional schematic structures of the FeFET and III-V/Si MOS phase shifter. The FeFET contains a thin $Hf_{0.5}Zr_{0.5}O_2$ film as the ferroelectric layer. The MOS phase shifter is formed by n-InGaAsP, $Al_2O_3$, and p-Si layers.

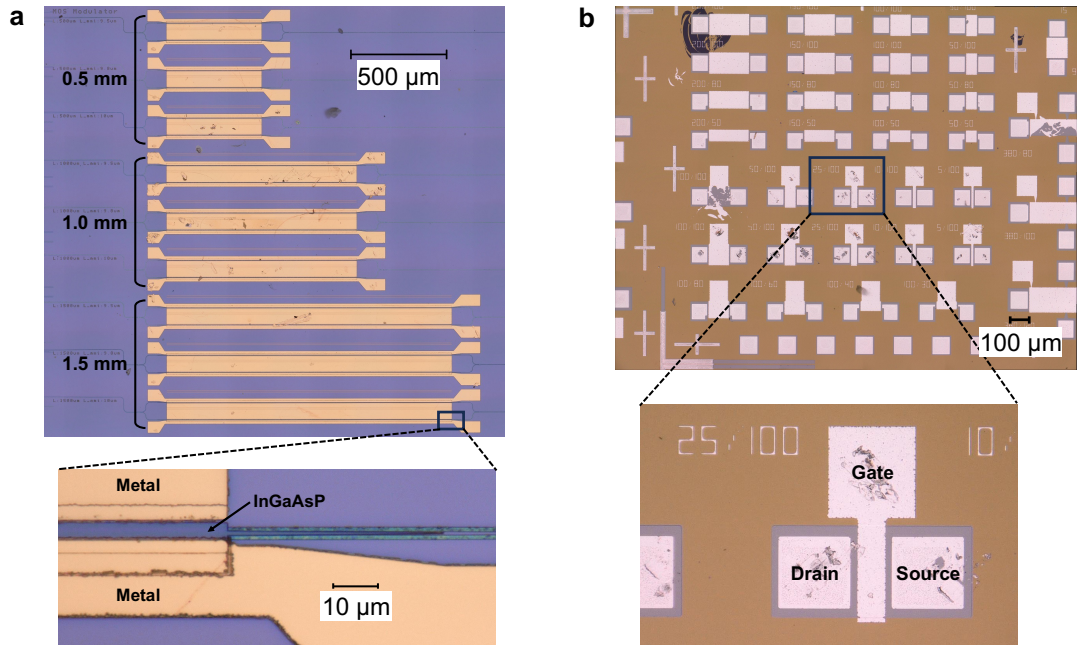

**Fig. 2. Fabricated devices. a**. Asymmetric Mach-Zehnder interferometers (AMZIs) with hybrid MOS phase shifters on the two arms. 3 pairs of AMZIs with various phase shifter lengths (0.5, 1.0, and 1.5 mm) are fabricated. **b**. FeFETs with various gate lengths and widths.

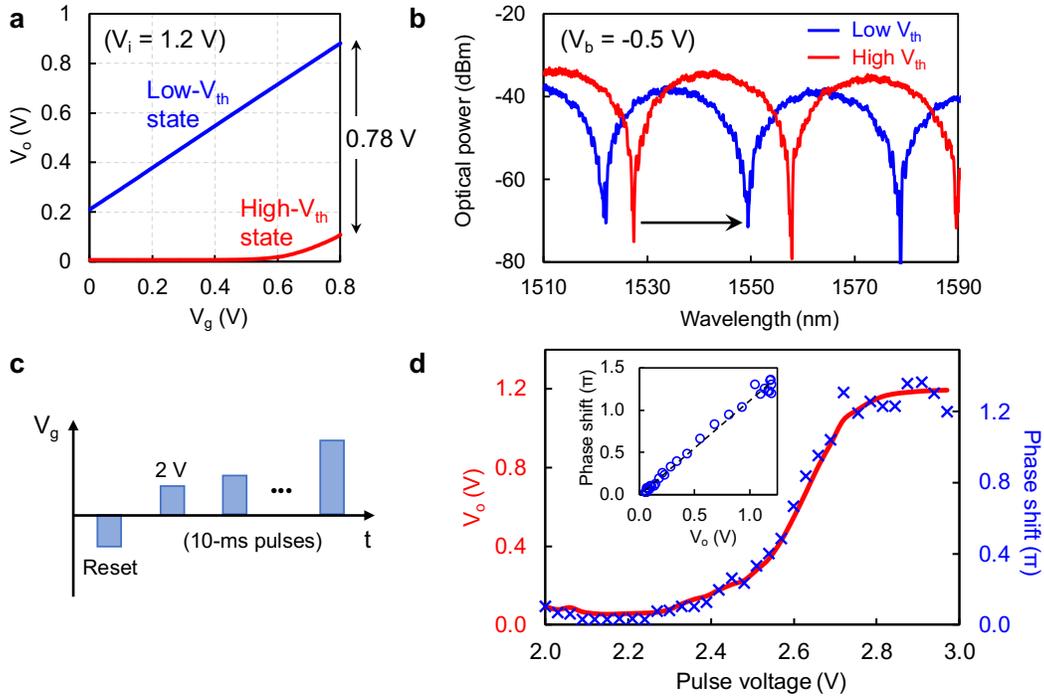

**Fig. 3. Non-volatile operations. a**. Source follower operations of a FeFET in two states with different threshold voltages. The threshold voltages are shifted by applying pulse voltages to the gate. **b**. Measured transmission spectra of the AMZI when the MOS phase shifter on one arm is driven by the FeFET. **c**. After an initial negative voltage pulse to reset the FeFET, positive voltage pulses from 2 to 3 V are applied to the gate sequentially. After each pulse, the output voltage of the FeFET ($V_o$) and the AMZI spectrum are measured. The phase shift is then extracted from the AMZI spectrum. **d**. Measured $V_o$ and the extracted phase shift as a function of the pulse voltage described in **c**. The inset shows the relationship between the phase shift and $V_o$.

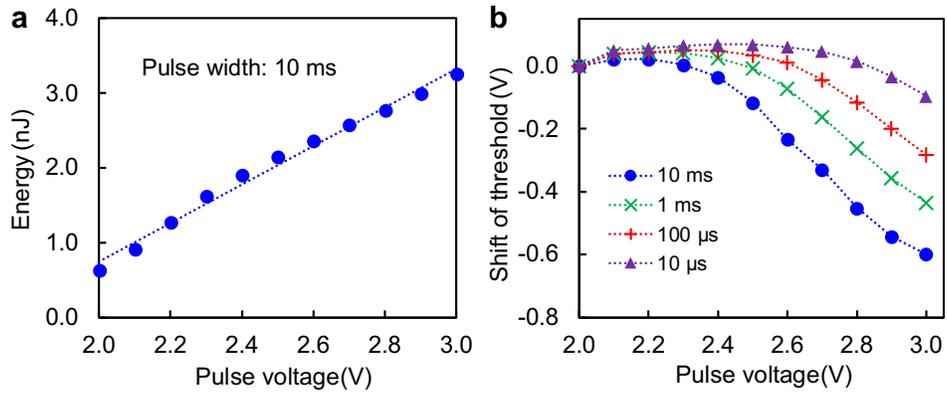

**Fig. 4. Switching characteristics. a**. Energy consumption of write-in operations using 10-ms pulses. The highest energy consumption is 3.2 nJ for a 3 V pulse. **b**. Shift of threshold voltage under various pulse widths. A short pulse induces a small amount of threshold shift compared with a long pulse with the same voltage.

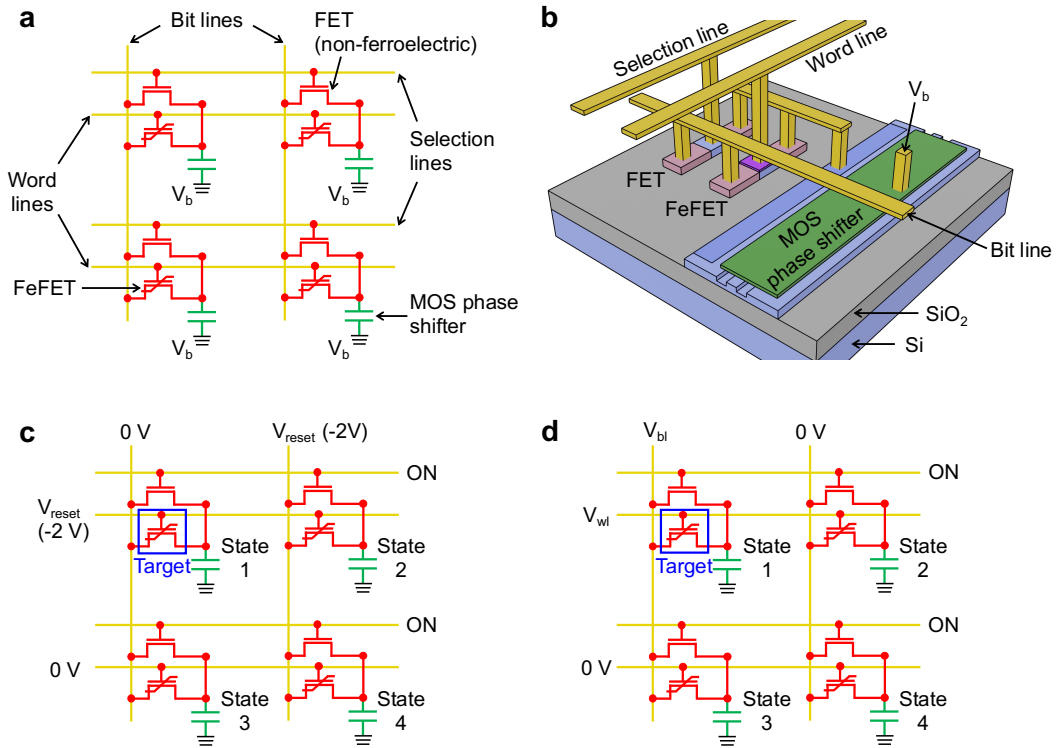

**Fig. 5. Crossbar array architecture. a**. Circuit diagram of the proposed "2T1C" crossbar array architecture, which consists of two FETs (one ferroelectric and the other non-ferroelectric) and one capacitor (the MOS phase shifter). **b**. Schematic structure of a single cell in the crossbar array. The transistors and optical waveguide are formed on the same Si layer. The drawing is not to scale. **c**. Voltage configurations for resetting a target FeFET in the crossbar array. The ON-state selection lines switch on the non-ferroelectric FETs so that the source and drain terminals of each FeFET have the same electric potential. **d**. Voltage configurations for programming a target FeFET in the crossbar array. $V_{wl}$ and $V_{bl}$ are chosen to only activate the target FeFET, without disturbing the rest.

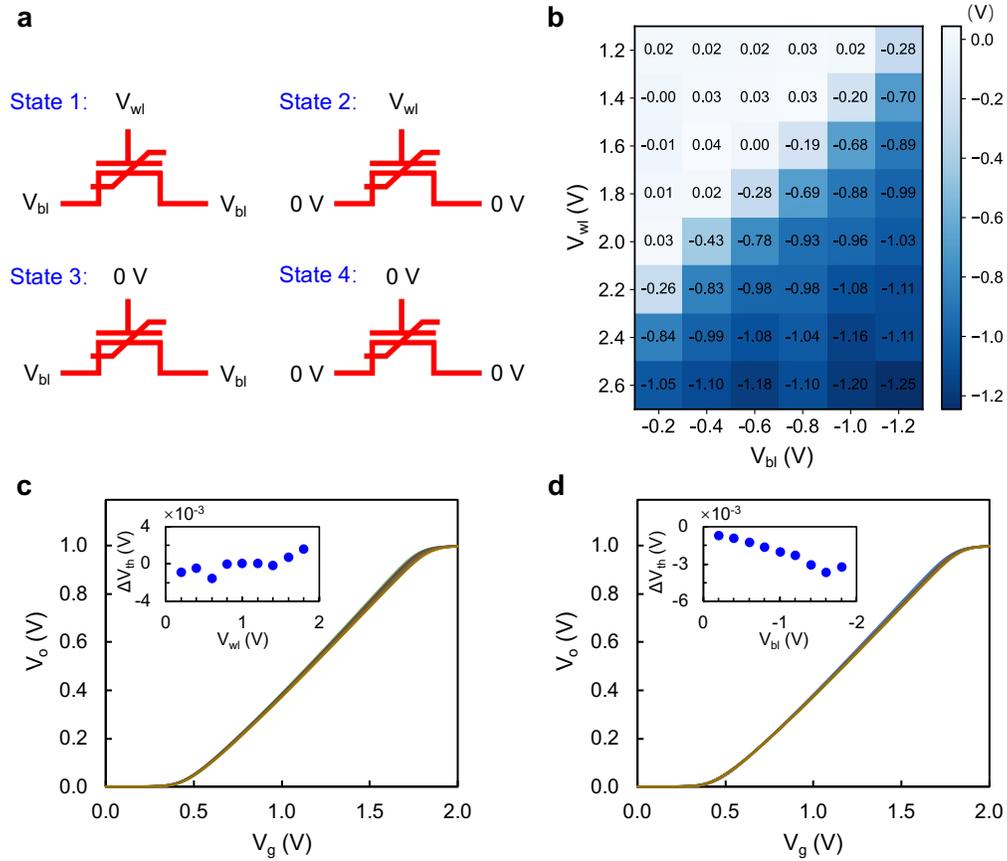

**Fig. 6. Responses of FeFETs in various states. a**. Four different states experienced by the FeFETs during write-in operations in the crossbar array. Ideally, only the target FeFET (state 1) is affected and all other FeFETs (states 2-4) should be unaffected. **b**. Shifts of the threshold voltage of a FeFET in state 1. After an initial pulse (2 V, 10 ms), the source follower operations of the FeFET under various combinations of $V_{wl}$ and $V_{bl}$ are characterized, from which the threshold voltages are extracted. **c-d**. Source follower operations of a FeFET in state 2 after application of various $V_{wl}$ pulses (**c**) and in state 3 after application of various $V_{bl}$ pulses (**d**), respectively. The insets show the shift of the extracted threshold voltage with respect to the initial state.

# Supplementary Information

# Non-volatile hybrid optical phase shifter driven by a ferroelectric transistor

Rui Tang[1,2], Kouhei Watanabe[1,2], Masahiro Fujita[1], Hanzhi Tang[1], Tomohiro Akazawa[1], Kasidit Toprasertpong[1], Shinichi Takagi[1], and Mitsuru Takenaka[1*]

[1]Department of Electrical Engineering and Information Systems, The University of Tokyo, Tokyo 113-8656, Japan
[2]These authors contributed equally
*Email: takenaka@mosfet.t.u-tokyo.ac.jp

## I. III-V/Si hybrid MOS phase shifter

The III-V/Si hybrid MOS phase shifter enables efficient and ultralow-power phase modulation with a CMOS-compatible voltage. Figure S1(a) shows the simulated mode profile in the phase shifter region. The MOS interface lies near the central region in order to obtain a high modulation efficiency. Figure S1(b) illustrates the taper that connects Si rib waveguide with III-V/Si hybrid waveguide. Due to a non-ideal taper design, the measured loss of the taper is 0.34 dB. In our subsequent design, we improved the taper design and successfully reduced the measured loss to almost zero. The measured propagation loss of MOS phase shifter is 1.26 dB/mm. This loss can also be significantly reduced since the entire Si layer in this chip is p-doped (doping concentration: $3\times10^{17}$ cm$^{-3}$). The characterized phase shift as a function of the applied voltage for the 1.5-mm-long phase shifter is shown in Fig. S1(c). The linear region (> 0.5 V) shows a slope of 1.94 π/V, corresponding to a $V_\pi L$ of 0.077 V·cm. The modulation efficiency can be further enhanced with improved designs. Previously, we have demonstrated a high modulation efficiency of 0.047 V·cm [1]. The dynamic modulation loss is 0.31 dB/π. The characterized power consumption of the phase shifter is shown in Fig. S1(d). The power consumption for π and 2π phase shifts are 13.3 pW and 96.3 pW, respectively.

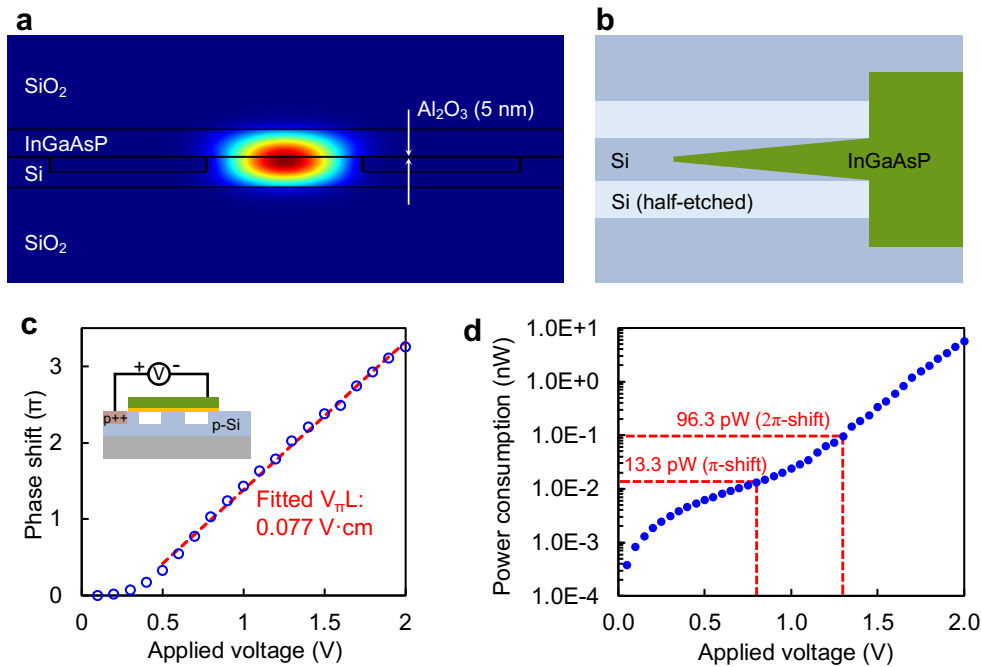

**Fig. S1**. **a**. Simulated mode profile in the phase shifter region. **b**. Top-view illustration of the taper that connects Si rib waveguide with III-V/Si hybrid waveguide. **c**. Characterized phase shift as a function of the applied voltage for the 1.5-mm-long phase shifter. **d**. Characterized power consumption of the phase shifter. The power consumption for $\pi$ and $2\pi$ phase shifts are 13.3 pW and 96.3 pW, respectively.

## II. HZO-based FeFET

FeFETs based on hafnium zirconium oxide (HZO) have promising applications in non-volatile memory and in-memory computation. Figure S2 shows the measured $I_d$-$V_g$ characteristics of an HZO-based FeFET with a gate length/width of 50/100 μm. The rising gate voltage induces the polarization change in the HZO layer and therefore causes the hysteresis in the $I_d$-$V_g$ curve.

We then consider the output voltage of a FET driving a capacitor in the source follower mode [2], as shown in Fig. 1(b). Under the condition of $V_{th} < V_g < V_i + V_{th}$, it is well known that the FET is working in the saturation regime and the current flowing between the drain and the source is given by

$$I_d = \frac{W}{2L}\mu_n C_{ox}(V_{gs} - V_{th})^2, \quad (S1)$$

where $W$ and $L$ are the channel width and length, respectively, $\mu_n$ is the effective electron mobility, $C_{ox}$ is the gate oxide capacitance per unit area, $V_{gs}$ is the voltage difference between the gate and the source [equals to $V_g$-$V_o$ in Fig. 1(b)]. Since the FET is driving a capacitor in our scheme, $I_{ds}$ is always 0 in the steady state. Therefore, we have $V_{gs} = V_{th}$, from which we obtain

$$V_o = V_g - V_{th}. \quad (S2)$$

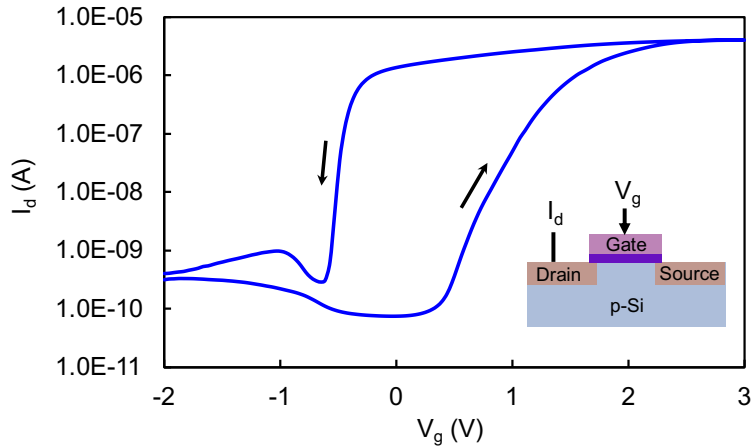

**Fig. S2**. $I_d$-$V_g$ characteristics of an HZO-based FeFET with a gate length/width of 50/100 μm.

Similar to the measurement described in Fig. 3(c), the electrical responses of multiple FeFETs with various gate lengths/widths to a series of gate pulses with increasing voltages are shown in Fig. S3. As can be seen, while the threshold voltages and memory windows are not exactly the same, electrical responses similar to that shown in Fig. 3(d) are obtained. A certain level of threshold shift is observed compared with the result in Fig. 3(d), which may be caused by the degradation of the MOS interface.

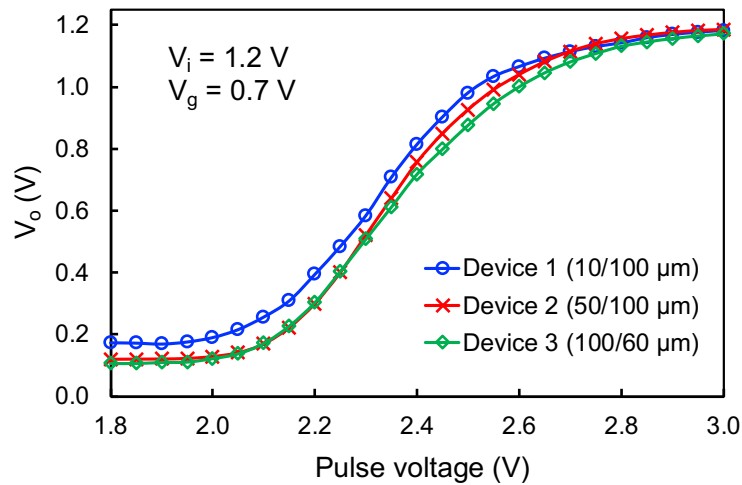

**Fig. S3**. Electrical responses of multiple FeFETs with various gate lengths/widths to a series of gate pulses with increasing voltages.